# PERFORMANCE IMPROVEMENT IN OFDM SYSTEM BY PAPR REDUCTION

Suverna Sengar[1], Partha Pratim Bhattacharya[2]

Department of Electronics and Communication Engineering,
Faculty of Engineering and Technology,
Mody Institute of Technology & Science (Deemed University),
Lakshmangarh, Dist – Sikar, Rajasthan, Pin – 332311, INDIA
[1]`suvernasengarmrt@gmail.com`, [2]`hereispartha@gmail.com`

## Abstract

*Orthogonal Frequency Division Multiplexing (OFDM) is an efficient method of data transmission for high speed communication systems. However, the main drawback of OFDM system is the high Peak to Average Power Ratio (PAPR) of the transmitted signals. OFDM consist of large number of independent subcarriers, as a result of which the amplitude of such a signal can have high peak values. Coding, phase rotation and clipping are among many PAPR reduction schemes that have been proposed to overcome this problem. Here two different PAPR reduction methods e.g. partial transmit sequence (PTS) and selective mapping (SLM) are used to reduce PAPR. Significant reduction in PAPR has been achieved using these techniques. The performances of the two methods are then compared.*

## 1. INTRODUCTION TO OFDM

Orthogonal frequency division multiplexing (OFDM) is a multicarrier modulation (MCM) technique which seems to be an attractive candidate for fourth generation (4G) wireless communication systems. OFDM offer high spectral efficiency, immune to the multipath delay, low inter-symbol interference (ISI), immunity to frequency selective fading and high power efficiency. Due to these merits OFDM is chosen as high data rate communication systems such as Digital Video Broadcasting (DVB) and based mobile worldwide interoperability for microwave access (mobile Wi-MAX). However OFDM system suffers from serious problem of high PAPR. In OFDM system output is superposition of multiple sub-carriers. In this case some instantaneous power output might increase greatly and become far higher than the mean power of system. To transmit signals with such high PAPR, it requires power amplifiers with very high power scope. These kinds of amplifiers are very expensive and have low efficiency-cost. If the peak power is too high, it could be out of the scope of the linear power amplifier. This gives rise to non-linear distortion which changes the superposition of the signal spectrum resulting in performance degradation. If no measure is taken to reduce the high PAPR, MIMO-OFDM system could face serious restriction for practical applications [1]-[4].

PAPR can be described by its complementary cumulative distribution function (CCDF). In this probabilistic approach certain schemes have been proposed by researchers. These include clipping, coding and signal scrambling techniques. Under the heading of signal scrambling techniques there are two schemes included. Which are Partial transmit sequence (PTS) and Selected Mapping (SLM). Although some techniques of PAPR reduction have been summarized in [5], it is still indeed needed to give a comprehensive review including some motivations of PAPR reductions, such as power saving, and to compare some typical methods of PAPR reduction through theoretical analysis and simulation results directly. An effective PAPR reduction technique should be given the best trade-off between the capacity of PAPR reduction and transmission power, data rate loss, implementation complexity and Bit-Error-Ratio (BER) performance etc.

In this paper, firstly the distribution of PAPR based on the characteristics of the OFDM signals are investigated then typical PAPR reduction techniques are analyzed.

## 2. OFDM SIGNAL CHARACTERISTICS

An OFDM symbol is made of sub-carriers modulated by constellations mapping. This mapping can be achieved from phase-shift keying (PSK) or quadrature amplitude modulation (QAM). For an OFDM system with N sub-carriers, the high-speed binary serial input stream is denoted as { $a_i$ }. After serial to parallel (S/P) conversion and constellation mapping, a new parallel signal sequence {$d_0,d_1,d_2,….d_i,…d_{N-1}$} is obtained, $d_i$ is a discrete complex-valued signal [6]. Here, $d_i \in$ {±1} when BPSK mapping is adopted. When QPSK mapping is used, $d_i \in$ {±1, ±i}.Each element of parallel signal sequence is supplied to N orthogonal sub-carriers {$e^{j2\pi f_0 t}, e^{j2\pi f_1 t}, ……. e^{j2\pi f_{N-1} t}$} for modulation, respectively. Finally, modulated signals are added together to form an OFDM symbol. Use of discrete Fourier transform simplifies the OFDM system structure. The complex envelope of the transmitted OFDM signals can be written as

$$x(t) = \frac{1}{\sqrt{N}} \sum_{k=0}^{N-1} X_k e^{j2\pi f_k t} , 0 \leq t \leq NT \qquad (1).$$

Signals with large N become Gaussian distributed with Probability Density Function (PDF) is given by [5].

$$P_r\{x(t)\} = \frac{1}{\sqrt{2\pi}\sigma} e^{-\frac{[x(t)]^2}{2\sigma^2}} \qquad (2)$$

where $\sigma$ is the variance of x(t).

### 2.1 PAPR

In general, the PAPR [3] of OFDM signals x(t) is defined as the ratio between the maximum instantaneous power and its average power

$$\text{PAPR}[\text{x(t)}] = \frac{P_{\text{PEAK}}}{P_{\text{AVERAGE}}} = 10 \log_{10} \frac{\max[|X(n)|^2]}{E[|x_n|^2]} \qquad (3)$$

where $P_{\text{PEAK}}$ represents peak output power, $P_{\text{AVERAGE}}$ means average output power. $E[\ \cdot\ ]$ denotes the expected value, $x_n$ represents the transmitted OFDM signals which are obtained by taking IFFT operation on modulated input symbols $X_k$ [7]. $x_n$ is expressed as:

$$x_n = \frac{1}{\sqrt{N}} \sum_{K=0}^{N-1} X_k W_N{}^{nk} \qquad (4).$$

The instantaneous output of an OFDM system often has large fluctuations compared to traditional single-carrier systems. This requires that system devices, such as power amplifiers, A/D converters and D/A converters, must have large linear dynamic ranges. If this is not satisfied, a series of undesirable interference is encountered when the peak signal goes into the non-linear region of devices at the transmitter, such as high out of band radiation and inter-modulation distortion. PAPR reduction techniques are therefore of great importance for OFDM systems. Also due to the large fluctuations in power output the HPA (high power amplifier) should have large dynamic range. This results in poor power efficiency.

## 3. PAPR REDUCTION TECHNIQUES

Several PAPR reduction techniques have been proposed in the literature [6]. These techniques are divided into two groups - signal scrambling techniques and signal distortion techniques which are given below:

**a) Signal Scrambling Techniques**
- Block Coding Techniques
- Block Coding Scheme with Error Correction
- Selected Mapping (SLM)
- Partial Transmit Sequence (PTS)
- Interleaving Technique
- Tone Reservation (TR)
- Tone Injection (TI)

**b) Signal Distortion Techniques**
- Peak Windowing
- Envelope Scaling
- Peak Reduction Carrier
- Clipping and Filtering

One of the most pragmatic and easiest approaches is clipping and filtering which can snip the signal at the transmitter is to eliminate the appearance of high peaks above a certain level. But due to non-linear distortion introduced by this process, orthogonality [8] is destroyed to some extent which results in In-band noise and Out-band noise. In-band noise cannot be removed by filtering, it decreases the bit error rate (BER). Out-band noise reduces the bandwidth efficiency but frequency domain filtering [7] can be employed to minimize the out-band power. Although filtering has a good effect on noise suppression, it may cause peak re-growth.

To overcome this drawback, the whole process is repeated several times until a desired situation is achieved. Here, two signal scrambling techniques are used to overcome these problems.

## 3.1. Signal Scrambling Techniques

The fundamental principle of these techniques is to scramble each OFDM signal with different scrambling sequences and select one which has the smallest PAPR value for transmission. Apparently, this technique does not guarantee reduction of PAPR value below to a certain threshold, but it can reduce the appearance probability of high PAPR to a great extent. This type of approach include: Selective Mapping (SLM) and Partial Transmit Sequences (PTS). SLM method applies scrambling rotation to all sub-carriers independently while PTS method only takes scrambling to part of the sub-carriers.

### 3.1.1 Selection Mapping Technique (SLM)

The CCDF of the original signal sequence PAPR above threshold PAPR0 is written as $P_r\{PAPR > PAPR_0\}$ . Thus for K statistical independent signal waveforms, CCDF can be written as $[P_r\{PAPR > PAPR_0\}]^R$, so the probability of PAPR exceed the same threshold. The probability of PAPR larger than a threshold Z can be written as

$$P(PAPR < Z) = F(Z)^N = (1 - \exp{(-Z)})^N \qquad (5).$$

Assuming that M-OFDM symbols carry the same information and that they are statistically independent of each other. In this case, the probability of PAPR greater than Z is equals to the product of each independent probability. This process can be written as

$$P\{PAPR_{LOW} > Z\} = (P\{PAPR > Z\})^M = ((1 - \exp{(-Z)})^N)^M \qquad (6).$$

In selection mapping method, firstly M statistically independent sequences which represent the same information are generated, and next, the resulting M statistically independent data blocks $S_m = [S_{m,0}, S_{m,1}, S_{m,N-1}]^T$, for m=1,2,...,M are then forwarded into IFFT operation simultaneously. $x_m = [x_1, x_2, x_N]^T$ in discrete time-domain are acquired and then the PAPR of these M vectors are calculated separately. Eventually, the sequences $x_d$ with the smallest PAPR is selected for final serial transmission. Figure 1 shows the basic block diagram of selection mapping technique for suppressing the high PAPR.

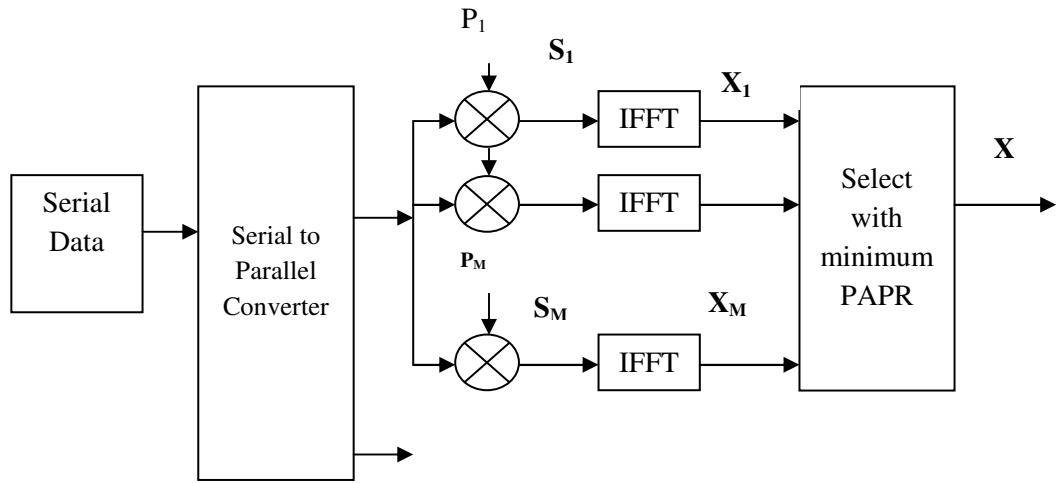

Figure 1. The Block Diagram of Selected Mapping Technique

### 3.1.2 Partial Transmit Sequence (PTS)

Partial Transmit Sequence (PTS) algorithm is a technique for improving the statistics of a multi-carrier signal. The basic idea of partial transmit sequences algorithm is to divide the original OFDM sequence [9] into several sub-sequences and for each sub-sequences multiplied by different weights until an optimum value is chosen.

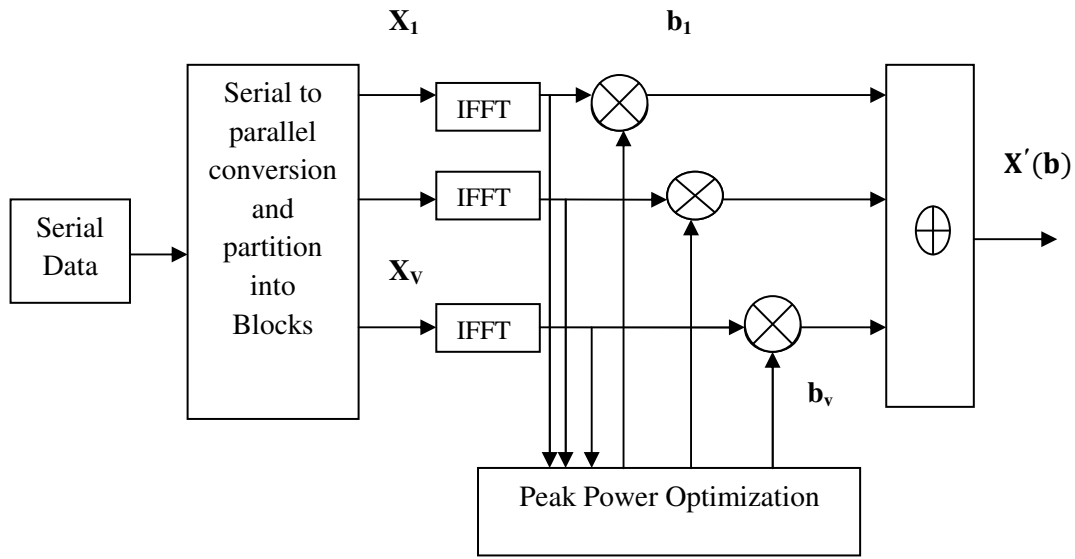

Figure 2. The Block diagram of PTS Technique

Figure 2 [10] is the block diagram of PTS technique. From the left side of diagram, the data information in frequency domain X is separated into V non-overlapping sub-blocks and each sub-block vectors has the same size N. So for each and every sub-block it contains N/V nonzero elements and set the rest part to zero. Assume that these sub-blocks have the same size and no gap between each other. The sub-block vector is given by

$$X = \sum_{v=1}^{V} b_v X_v \qquad (7)$$

where $b_v = e^{j\varphi_v}(\varphi_v \in [0, 2\pi])\{v = 1, 2, ..., X_v\}$ is a weighting factor been used for phase rotation. The signal in time domain is obtained by applying IFFT operation [11] on, that is

$$\hat{x} = IFFT(X) = \sum_{v=1}^{V} b_v IFFT(X_v) = \sum_{v=1}^{V} b_v X_v \qquad (8).$$

For the optimum result one of the suitable factor from combination b = [b₁, b₂,.., bᵥ] is selected and the combination is given by

$$b = [b_1, b_2, ..., b_v] = \arg\min_{(b_1, b_2, ..., b_v)}(max_{1 \le n \le N}|\sum_{v=1}^{V} b_v X_v|^2) \qquad (9)$$

where arg min [(·)] is the condition that minimize the output value of function.

## 4. REDUCTION OF PAPR

SLM and PTS algorithms are two typical non-distortion techniques for reduction of PAPR in OFDM system [12]-[16]. SLM method [15] applies scrambling rotation to all sub-carriers independently while PTS methods [16] only take scrambling to part of the sub-carrier.

Table 1 Parameters used in SLM and PTS algorithm

| Parameters | Values used |
|---|---|
| Number of sub-carriers (N) | 64, 128 |
| Oversampling factor (OF) | 8 |
| Modulation scheme | QPSK |
| Route numbers used in SLM method (M) | 2, 4, 8, 16 |
| Number of sub-blocks used in PTS methods (V) | 4 |
| Total number of combinations or IFFT for weighting factor 1 and 2 | 16, 256 |
| Number of generated OFDM signal | 1000 |

Table 1 shows the parameters of OFDM signal which is used for PAPR reduction. Here, the number of sub-carriers used are N=64, 128 and the pseudo-random partition scheme is applied for each carrier, adopting QPSK constellation mapping, weighting factor being $b_v \in [\pm 1, \pm j]$. The flow chart used for PAPR reduction technique is given in Figure 3.

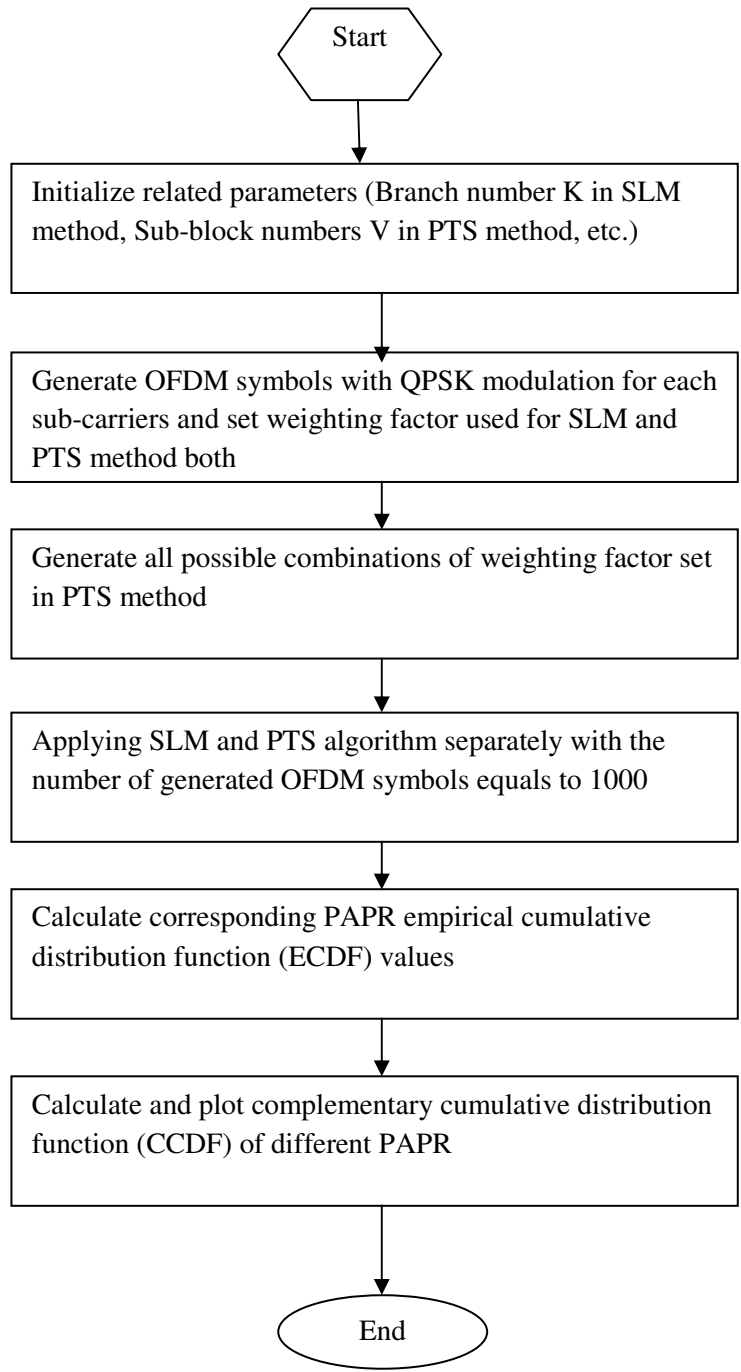

Figure 3. Flow Chart of PAPR Reduction Performances in SLM and PTS Methods

# 5. SIMULATION RESULTS

Figure 4 shows the CCDF as a function of PAPR distribution when SLM method is used with 64 numbers of subcarrier. Figure 5 shows the same result for 128 numbers of subcarrier. M takes the value of 1 (without adopting SLM method), 2, 4, 8 and 16. It is seen in Figure 4 and Figure 5 that with increase of branch number M, PAPR's CCDF gets smaller.

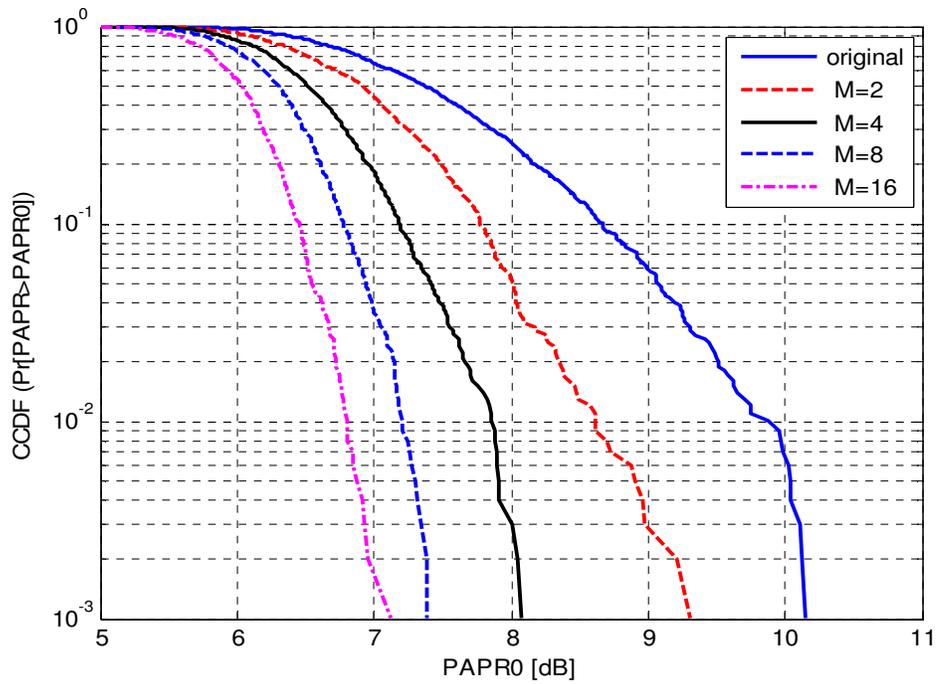

Figure 4. PAPR's CCDF using SLM method with N=64

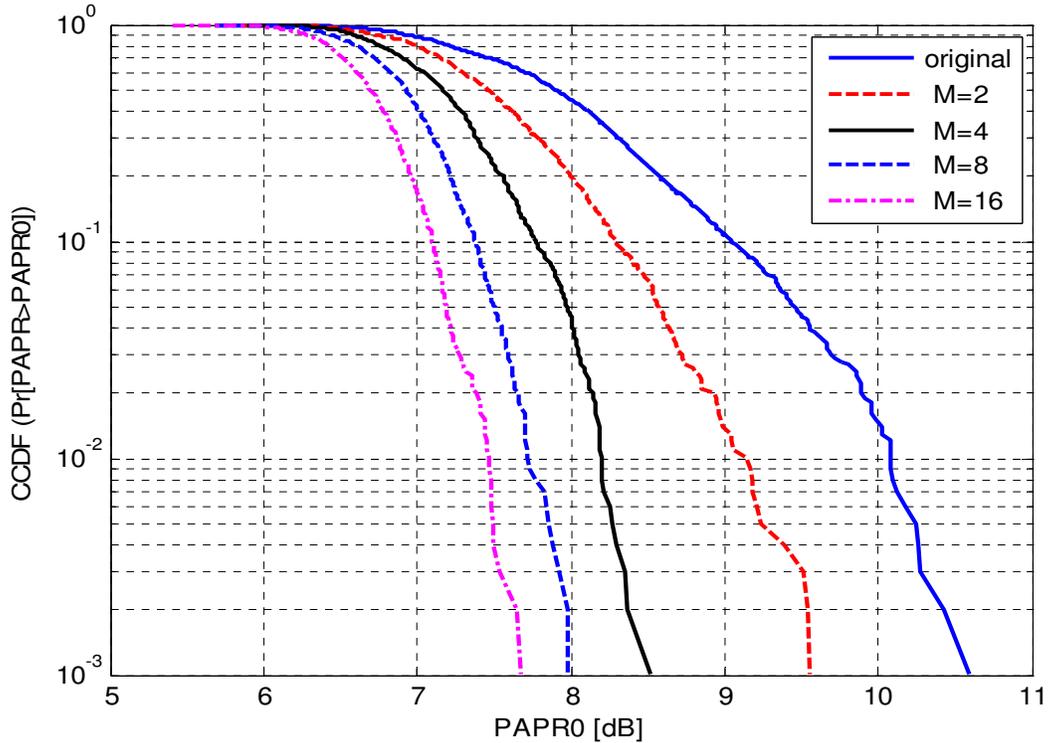

Figure 5. PAPR's CCDF using SLM method with N=128

Now discussed the simulation result for PTS technique, there are varying parameters which impact the PAPR reduction performance these are: 1) The number of sub-blocks V, which influences the complexity strongly; 2) The number of possible phase value W, which impacts the complexity; and 3) The sub-block partition schemes. Here, only one parameter is considered that is sub-block size V.

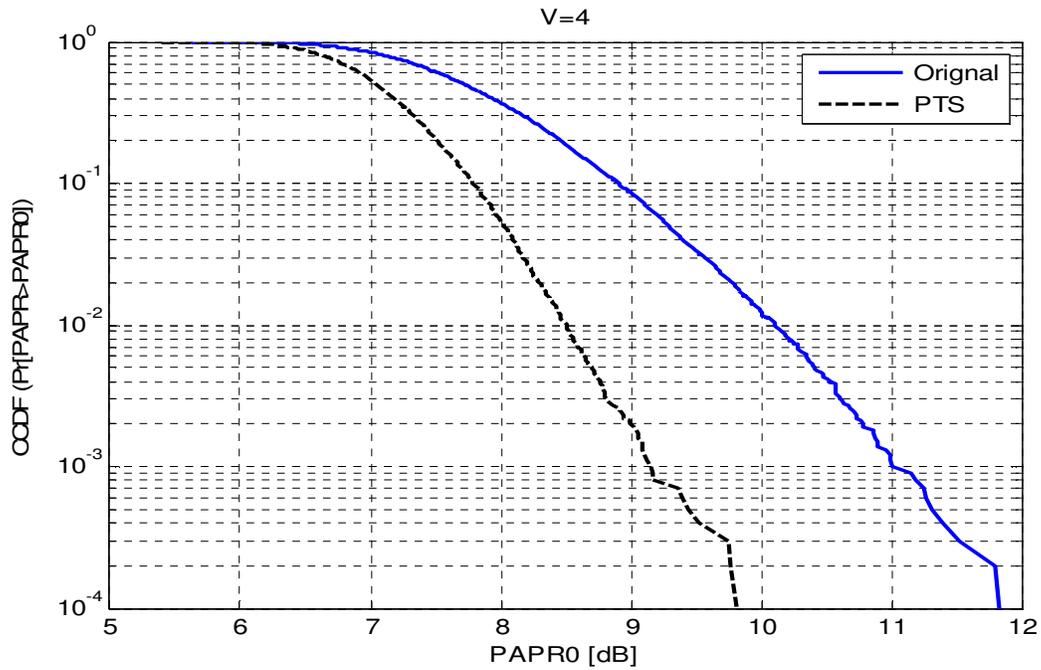

Figure 6. PAPR's CCDF using PTS method with N=256 and V=4

Figure 6 shows that PTS technique improves the performance of OFDM system, moreover, it can be shown that with increasing the value of V the PAPR performance becomes better.

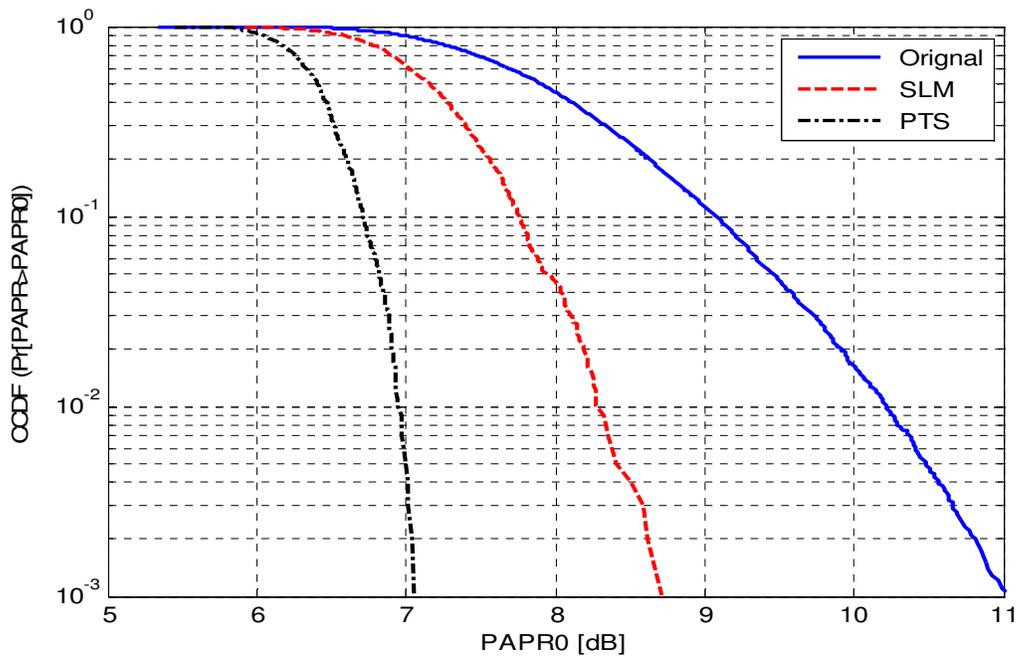

Figure 7. PAPR's CCDF using SLM and PTS method with N=64

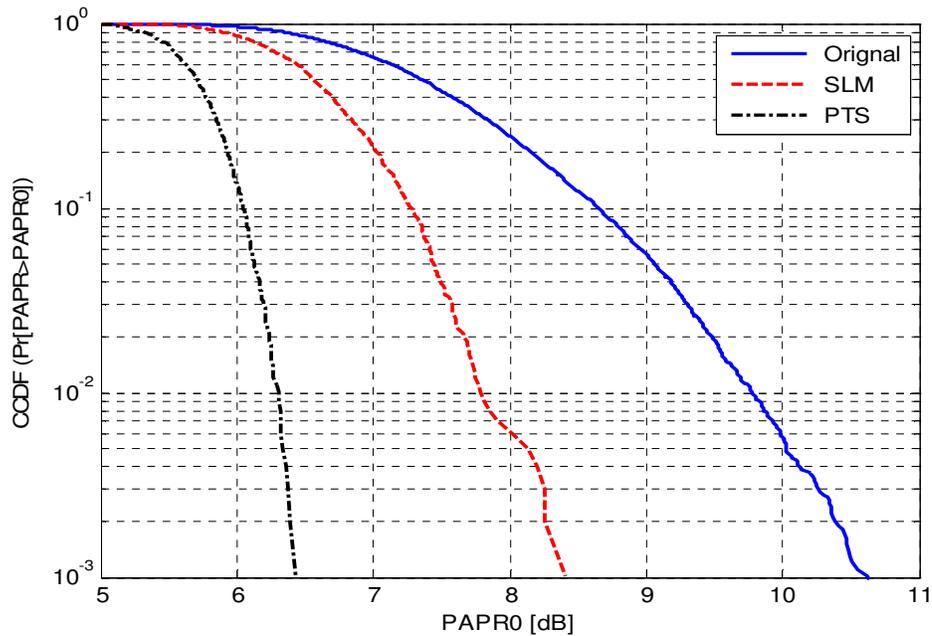

Figure 8. PAPR's CCDF using SLM and PTS method with N=128

In Figure 7 and Figure 8 it is clear that PTS method provides a better PAPR reduction performance compared to SLM method.

## 6. CONCLUSIONS

OFDM is a very attractive technique for wireless communications due to its spectrum efficiency and channel robustness. One of the serious drawbacks of OFDM systems is that the composite transmit signal can exhibit a very high PAPR when the input sequences are highly correlated. In this paper, several important aspects are described as well as mathematical analysis is provided, including the distribution of the PAPR used in OFDM systems. Two typical signal scrambling techniques, SLM and PTS are investigated to reduce PAPR, all of which have the potential to provide substantial reduction in PAPR. PTS method performs better than SLM method in reducing PAPR.

## 7. ACKNOWLEDGEMENTS

The author would like to thank Mody Institute of Technology & Science for supporting the work.

**Authors**

Suverna Sengar was born in India on September 25, 1990. She received B.Tech (2010) in Electronics & Instrumentation from Uttar Pradesh Technical University (U.P.T.U), Lucknow. She is a final year student of M.Tech in Signal Processing, Mody Institute of Technology and Science (Deemed University), Rajasthan, India.

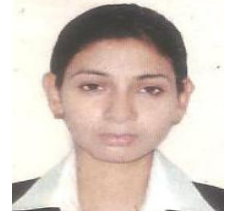

Partha Pratim Bhattacharya was born in India on January 3, 1971. He did M. Sc in Electronic Science from Calcutta University, India in 1994, M. Tech from Burdwan University, India in 1997 and Ph.D (Engg.) from Jadavpur University, India in 2007.

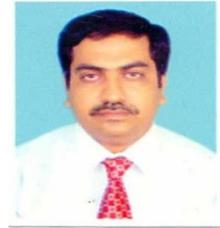

He has 15 years of experience in teaching and research. He served many reputed educational Institutes in India in various positions starting from Lecturer to Professor and Principal. At present he is working as Professor in Department of Electronics and Communication Engineering in the Faculty of Engineering and Technology, Mody Institute of Technology and Science (Deemed University), Rajasthan, India. He worked on Microwave devices and systems and mobile cellular communication systems. He has published 60 papers in refereed journals and conferences. His broad research interest includes wireless communication.

Dr. Bhattacharya is a member of The Institution of Electronics and Telecommunication Engineers, India and The Institution of Engineers, India. He received Young Scientist Award from International Union of Radio Science in 2005. He is working as the editorial board member and reviewer in many reputed journals.